\documentclass[iop]{emulateapj}
\usepackage {graphicx}

\begin{document}

\submitted{Accepted to ApJ, May 4, 2016}

\title{Star Cluster Formation and Destruction in the Merging Galaxy NGC 3256}

\author{A. J. Mulia\altaffilmark{1},
R. Chandar\altaffilmark{1},
B. C. Whitmore\altaffilmark{2}}

\altaffiltext{1}{Physics \& Astronomy Department, University of Toledo, Toledo, OH 43606-3390}
\altaffiltext{2}{Space Telescope Science Institute, 3700 San Martin Drive, Baltimore, MD 21218}

\vspace{.2in}

\begin{abstract}
We use the Advanced Camera for Surveys on the \emph{Hubble Space
  Telescope} to study the rich population of young massive star clusters in
the main body of NGC 3256, a merging pair of galaxies with a high star
formation rate (SFR) and SFR per unit area ($\Sigma_{\rm{SFR}}$). These clusters
have luminosity and mass functions that follow power laws, $dN/dL \propto
L^{\alpha}$ with $\alpha = -2.23 \pm 0.07$, and $dN/dM \propto
M^{\beta}$ with $\beta = -1.86 \pm 0.34$ for $\tau < 10$ Myr clusters, similar to
those found in more quiescent galaxies. The age distribution
can be described by $dN/d\tau \propto \tau ^ \gamma$, with $\gamma
\approx -0.67 \pm 0.08$ for clusters younger than about a few hundred
million years, with no obvious dependence on cluster mass. This is consistent
with a picture where $\sim 80 \%$ of the clusters are
disrupted each decade in time. We investigate the claim that galaxies with high
$\Sigma_{\rm{SFR}}$ form clusters more efficiently than quiescent
systems by determining the fraction of stars in bound clusters
($\Gamma$) and the CMF/SFR statistic (CMF is the cluster mass
function) for NGC 3256 and comparing the
results with those for other galaxies. We find that the
CMF/SFR statistic for NGC 3256 agrees well with that found for
galaxies with $\Sigma_{\rm{SFR}}$ and SFRs that are lower by $1-3$
orders of magnitude, but that estimates for $\Gamma$ are only robust
when the same sets of assumptions are applied. Currently, $\Gamma$
values available in the literature have used different sets of
assumptions, making it more difficult to compare the results between galaxies.
\end{abstract}


\section{INTRODUCTION}

NGC 3256 is a merging pair of galaxies $\approx 36$ Mpc away. Dynamical
simulations suggest that the system began interacting $\approx 450$
Myr ago, and it has since undergone a period of major star and star
cluster formation. It exhibits two tidal tails that are rich
with young massive stellar clusters (Knierman et al. 2003; Maybhate et
al. 2007; Trancho et al. 2007a; Mulia, Chandar, \& Whitmore 2015). The
main body of NGC 3256 contains a dense population of
clusters, many of which are embedded in the galaxy's dusty
interstellar medium. The galaxy's intense star formation makes it an
extreme environment in which to study cluster formation and destruction.

The cluster population of NGC 3256 has been studied in a number of
previous works. Zepf et al. (1999) used $B$ and $I$ band images taken
with the Wide Field Planetary Camera 2 on the \emph{Hubble Space
  Telescope} (\emph{HST}) to examine the colors and luminosities of
main body clusters. Using the fraction of blue light that they found
in clusters, they estimated that the efficiency of cluster formation
in NGC 3256 is $\sim 20\%$. Trancho et al. (2007b) performed optical
spectroscopy of 23 clusters in the main body of NGC 3256, finding ages
of a few to $\sim 150$ Myr and masses from $(2-40) \times 10^5$
$M_{\odot}$. Goddard, Bastian, \& Kennicutt (2010; hereafter G10)
estimated ages and masses of NGC 3256 clusters from $UBVI$ photometry
using \emph{HST}'s Advanced Camera for Surveys (ACS) and estimated the
fraction of stars forming in bound clusters, hereafter referred to as
$\Gamma$, to be $\Gamma = 22.9\% ^{+7.3}_{-9.8}$. Their method
involved estimating the cluster formation rate (CFR), taken from the total
mass of clusters younger than 10 Myr, and dividing by the galaxy's
star formation rate (SFR).

Some works, including G10, have suggested that there is a
  correlation between $\Gamma$ and SFR density,
$\Sigma_{\rm{SFR}}$ (e.g., Silva-Villa \& Larsen 2011; Cook et
al. 2012), and between $T_{L}(U)$ (the fraction of $U$ band light
  contained in clusters) and $\Sigma_{\rm{SFR}}$ (Larsen
  \& Richtler 2000; Adamo \"{O}stlin, \& Zackrisson 2011). These
  relationships imply that the conditions for
cluster formation are not universal, but are dependent on
conditions in the host galaxy. Chandar, Fall, \& Whitmore
(2015), on the other hand, measured
the cluster mass function (CMF) in seven galaxies and found that when
normalized by the SFR, these CMFs fall nearly on top of one another. The
CMF/SFR statistic implies that cluster formation is
similar in many galaxies, regardless of SFR. We estimate both $\Gamma$
and, for the first time, the CMF/SFR statistic, in NGC 3256.

This paper is arranged as follows. Section \ref{obs} describes the
observations and cluster selection. Section \ref{method} describes the method
for obtaining ages and masses of clusters. Section \ref{results} presents the
luminosity functions (LFs), age distributions, and mass functions for NGC
3256. In Section \ref{dist}, we quantify the effects that distance has on
the measured LF and age distribution of the clusters. We
determine the CMF/SFR statistic and $\Gamma$ from our new NGC 3256
cluster sample in Section \ref{discussion}. We summarize our results
and state conclusions in Section \ref{conc}.

\section{OBSERVATIONS, DATA REDUCTION, AND CLUSTER SELECTION}
\label{obs}

\subsection{Observations and Data Reduction}
\label{obs_red}

Our observations come from the ACS on \emph{HST}. NGC 3256 was
observed using the filters $F555W$ ($\approx V$ in the
Johnson--Cousins system; exposed for 2552 s), $FR656N$ (H$\alpha$;
2552 s), and $F330W$ ($\approx U$; 11358 s) as part of the program GO-9735 (PI:
Whitmore). The $V$ and $U$ band images were taken in 2003 November using the
Wide Field Camera (WFC) and High Resolution Camera (HRC), respectively.
The H$\alpha$ observations were taken in 2004 March. WFC
observations using $F435W$ ($\approx B$) and $F814W$
($\approx I$) filters were taken in 2005 November as part of program
GO-10592 (PI: Evans) for 1320 and 760 s, respectively.

The raw data were processed through the standard ACS pipeline.
The reduced, multidrizzled WFC images were taken from the Hubble Legacy
Archive (HLA) and have a scale of 0.05$''$ per pixel, while $U$ band images
taken with HRC have 0.025$''$ per pixel. A $BVI$H$\alpha$ color image of
NGC 3256 is shown in Figure \ref{image}.

\subsection{Cluster Selection}
\label{clus_sel}

We use the IRAF\footnote{IRAF is distributed by the National Optical
  Astronomy Observatory, which is operated by the Association of
  Universities for Research in Astronomy (AURA) under a cooperative
  agreement with the National Science Foundation.} task DAOFIND on the
$I$ band image to make an initial catalog of point sources in NGC
3256. Using the PHOT task in IRAF, we run photometry
on the $B$, $V$, and $I$ band images using an aperture of 3 pixels in radius and
a background area between 5 and 8 pixels in radius. For the $U$ band image, we
use an aperture radius of 6 pixels and a background area of radius
$10-16$ pixels, since the HRC has twice the resolution of the WFC. We
use the ACS photometric zeropoint calculator\footnote{http://www.stsci.edu/hst/acs/analysis/zeropoints}
to convert instrumental magnitudes to the VEGAMAG system.  We use
aperture corrections from 3 pixels to infinity in $B$, $V$, and $I$ bands taken from isolated
star clusters found in the tidal tails of NGC 3256 and reported in Mulia,
Chandar, and Whitmore (2015). For the $U$ band aperture correction, we
find $\sim$ 15 isolated bright clusters and measure their mean 6 -- 20
pixel magnitude.  We use the encircled energy catalog found in Table
4 of Sirianni et al. (2005) for the aperture correction from 20
pixels to infinity. The final corrections were 0.63, 0.45, 0.42, and 0.54 mag
for $U$, $B$, $V$, and $I$ band filters respectively.  We also correct the
photometry in each filter for foreground extinction, 0.528, 0.441,
0.334, 0.264, and 0.183 mag for $U$, $B$, $V$, H$\alpha$, and $I$,
respectively, taken from the NASA/IPAC Extragalactic Database.

We require photometry in all five $UBVI$H$\alpha$ bands in order to
accurately age-date star clusters. This has a minimal impact on our
catalog and removes only a small fraction ($<10\%$) of detected
sources. We visually inspect all sources to remove cosmic rays and hot
pixels from the catalog.

The low galactic latitude of NGC 3256 means that there will be a
significant number of foreground stars from the Milky Way in our images.
Luckily, at the distance of NGC 3256, we expect most clusters to be at
least partially resolved in the \emph{HST}/WFC images. We measure the
concentration index, $C$, to aid our separation of foreground stars
and cluster candidates. For a single source, $C$ is measured as the
difference in magnitude taken at two different apertures
(typically done in the $V$ band at 0.5 and 3.0 pixels, which we use
here). $C$ has proven to be a robust method in determining the
compactness of an object and has therefore often been used to remove
stellar contamination from cluster catalogs in many previous works
(see Chandar et al. 2010 for a more in-depth discussion on
$C$). Figure \ref{CI} shows $m_V$ versus $C$ for all cluster
candidates (circles) and likely foreground stars (asterisks). All
stars fall to the left of the vertical dashed line at $C=2.2$. We
therefore only include cluster candidates with $C > 2.2$.

\begin{figure}[ht]
\centering
\includegraphics[scale=0.2]{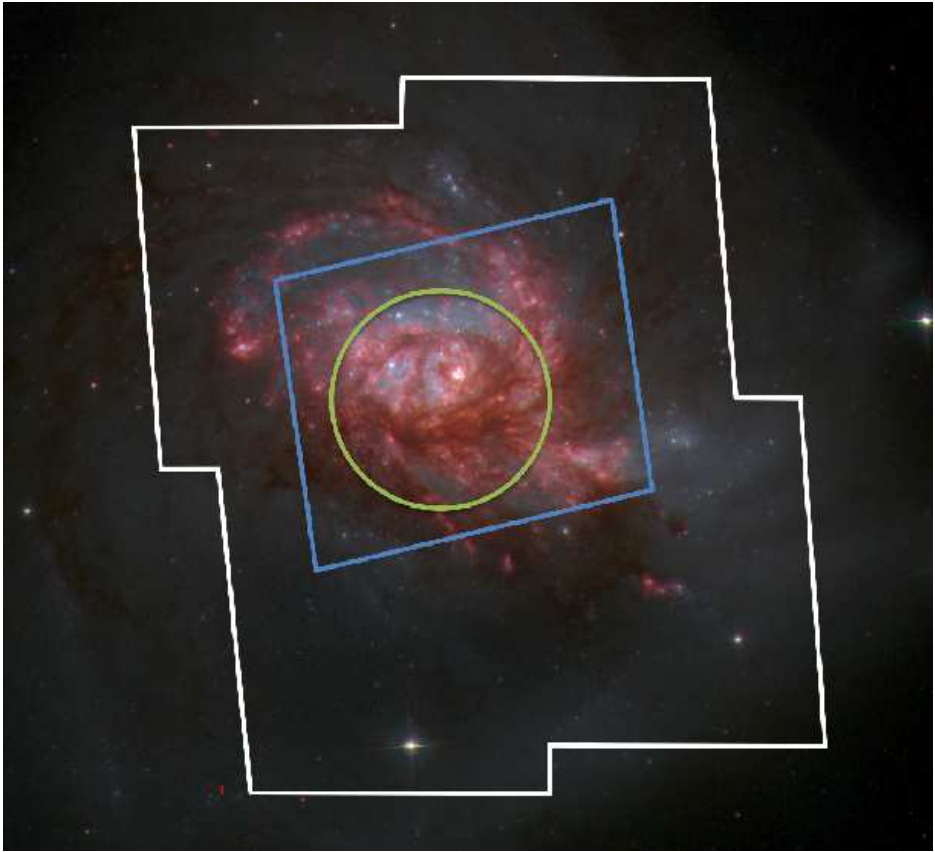}
\caption{$BVI$ color image of NGC 3256, with H$\alpha$ overlaid. The
  green circle indicates where we divide our sample into inner and outer
  regions and is $\approx 9.5''$ (2 kpc) in radius. The area covered
  by G10 is shown as the blue rectangle and covers $\approx 3$ times
  less area than our coverage, shown as the white perimeter.}
\label{image}
\end{figure}

\begin{figure}[ht]
\centering
\includegraphics[scale=1.0]{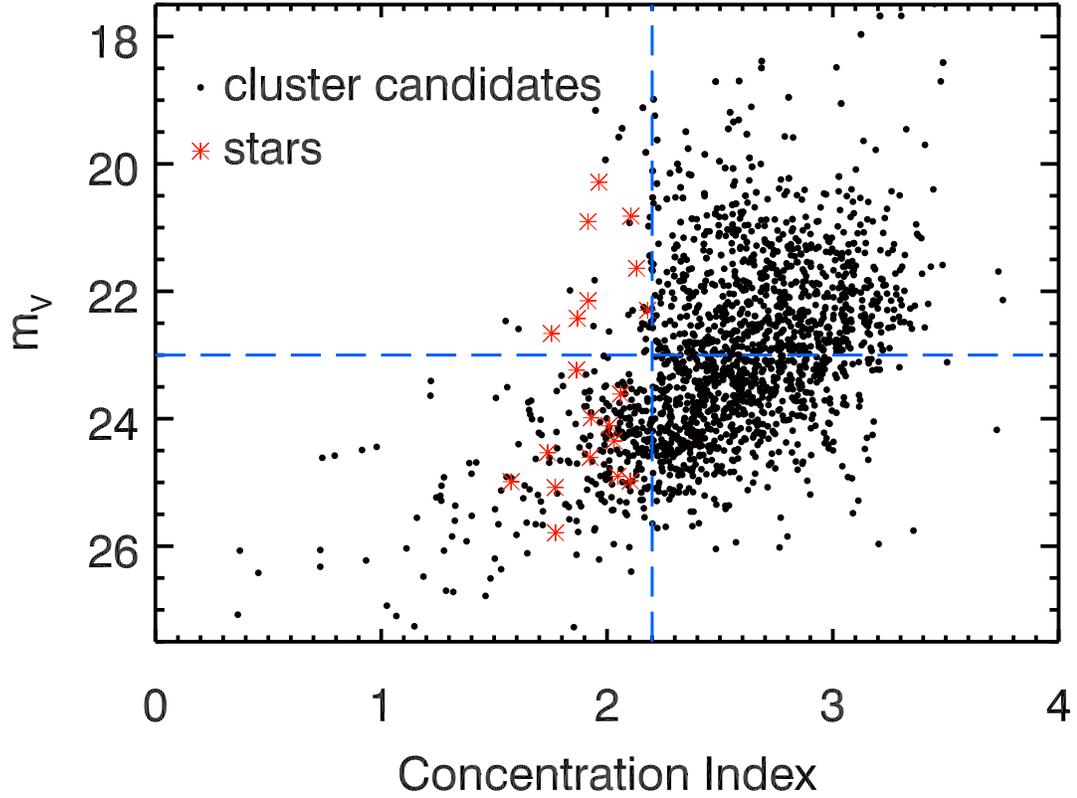}
\vspace{-3.00mm}
\caption{Plot of concentration index vs. brightness for all sources found in
  NGC 3256. The blue vertical line at $C=2.2$ illustrates that all likely
  field stars are more centrally concentrated than $C=2.2$. We note
  that the apparent relation between concentration index and
  brightness is an artifact of incompleteness and is not present above
  the horizontal line that illustrates our cutoff of $m_V = 23.0$}
\label{CI}
\end{figure}

We address incompleteness in our catalog by making a cut where the
$m_V$ distribution begins to flatten at the faint end, rather than
continue in a power law fashion. Because the central
portion of NGC 3256 is dustier and has a higher background level than
the outer regions, we use separate completeness limits for the inner
and outer regions. Clusters in the inner region are complete down to
$m_V \approx 21.5$, while the outer region is complete down to $m_V \approx
23.0$. The separation of these two regions is shown in Figure \ref{image}.
We find that contamination from background galaxies is negligible
in the main body of NGC 3256. In all, we find 505 cluster candidates
that meet our selection criteria, and compile their properties in
Table \ref{catalog}.

We briefly compare our catalog to that of G10, who studied clusters in
the central portion of NGC 3256 using $UVI$ filters taken with the
HRC (see Figure \ref{image}) and the $B$ band taken with the WFC
from \emph{HST}/ACS. Their final sample covered the central $5.8 \times
5.8$ kpc$^2$ and included 276 clusters after imposing various
selection criteria on photometric uncertainty
and on the goodness of fit to simple stellar population (SSP)
models. The portion of NGC 3256 studied here extends the G10 coverage
to include the northern spiral arm, as well
as outside the central, brightest portion of the
system (as shown in Figure \ref{image}). In addition, we include
H$\alpha$ observations in our analysis, which can be
critical for breaking the degeneracy between age and extinction, a degeneracy
that affects a number of clusters in this merging system (see Section
\ref{ha_calib} for more details). Our catalog includes more
  than $95\%$ of the sources that were found by G10, when both
  catalogs are matched in area and magnitude ($m_V < 21.5$). Conversely, G10
  only find $\sim 70\%$ of the clusters in our catalog (again when
  matched in area and magnitude). Our study covers an area $\approx$ 3.5
  times larger than that of G10, although we note a lack
  of clusters south of the main body.

\section{Cluster Age and Mass Determination}
\label{method}

\begin{figure}[ht]
\begin{center}
\includegraphics[scale=1.0]{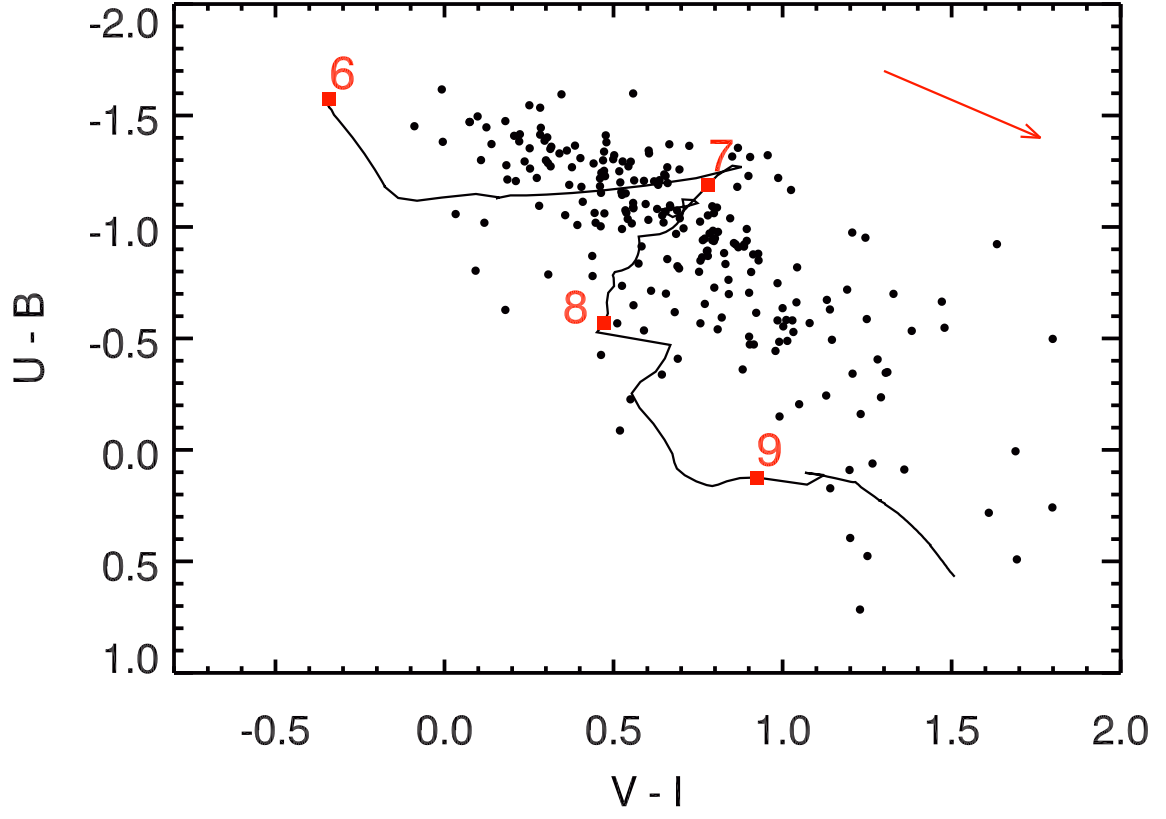}
\vspace{-4.00mm}
\caption{Color-color diagram of main body clusters in NGC 3256.  For
  clarity, we show only the brightest clusters ($m_V < 21.5$). The red arrow
  shows the reddening vector for $A_V=1$. Various ages in
  log$(\tau/\rm{yr})$ are shown by numbered red squares.}
\label{CCD}
\end{center}
\end{figure}

We present the color-color diagram for star clusters
in NGC 3256 in Figure \ref{CCD}. The solid line is a stellar population model
from G. Bruzual \& S. Charlot (2006, private communication, hereafter
BC06; and see also Bruzual \& Charlot 2003) that predicts the
evolution of star clusters from about $10^6$ -- $10^{10}$ yr.
All models assume a metallicity of Z = 0.02 (i.e., solar metallicity).
Numbers mark the logarithmic age ($\tau$) corresponding to the population.

\subsection{Spectral Energy Distribution (SED) Fitting}
\label{fitting}

Using $UBVI$H$\alpha$ photometry, we fit photometric SEDs of clusters
to SEDs from cluster evolution tracks of BC06 in order to determine
the age of each cluster. Reddening due to dust can yield inaccurate
ages if not accounted for, because redder
colors imply older ages.  We therefore adopt the common technique of
varying age and $E(B-V)$ in order to minimize the function

\begin{equation}
\chi^{2}(\tau,E(B-V)) = \sum\limits_{\lambda} W_\lambda
(m^{\mathrm{obs}}_\lambda - m^{\mathrm{mod}}_\lambda)^2.
\end{equation}

\noindent Here, $m^{\mathrm{obs}}_\lambda$ is the observed magnitude and has been
corrected for both aperture and foreground extinction
(see Section \ref{clus_sel} for details), and the BC06 models
($m^{\mathrm{mod}}_{\lambda}$) are normalized to each
cluster's apparent $V$ band magnitude. The function
$W_\lambda=(\sigma^{2}_\lambda + 0.05^2)^{-1}$ weights each
photometric measurement in the fit by accounting for the
photometric uncertainty ($\sigma_{\lambda}$) output by PHOT.  We add
0.05 in quadrature to $\sigma_{\lambda}$ to ensure that no single measurement
dominates the SED fit. 

The BC06 model predicts a mass-to-light ratio (for the $V$ band
luminosity) for each age, which we use to estimate a mass for each
cluster. $L_V$ is calculated from each cluster's $M_V$ magnitude
(corrected for internal extinction) and the distance modulus of NGC
3256. The masses are computed using a Chabrier (2003) initial mass
function.

\subsection{Calibrating the H$\alpha$ Filter}
\label{ha_calib}

Broad-band colors alone make it difficult to differentiate clusters
that are red because they are older from those that are young but
appear redder due to the presence of dust. This is known as the
age-extinction degeneracy. Including H$\alpha$ photometry as a fifth
point in the SED can help to break this degeneracy
and has been used in a number of studies to age-date clusters
(e.g., Fall, Chandar, \& Whitmore 2005; Chandar et
al. 2010). In this section we describe our method for calibrating the
H$\alpha$ filter in order to use it in the SED fitting method.

We first determine the zero point of the $FR656N$ filter. We identify
$\approx 20$ somewhat older star clusters, those that have only
continuum emission from stars (i.e., no line emission from ionized
gas). We estimate the age and extinction of each of these line-free
clusters by performing the $\chi^2$ fit described in Section
\ref{fitting} using only the $UBVI$ filters to predict a magnitude in
the $FR656N$ band. Comparison between the predicted and instrumental
magnitudes gives a zero point of $22.4 \pm 0.1$. We find that shifting
the zero point by 0.1 alters the ages of only $\sim 5\%$ of the
clusters by $\ge$ 0.3 in log($\tau/\rm{yr}$).

Next, we assess whether or not we need to make an adjustment to the
predictions for H$\alpha$ line emission due to the escape of ionizing
photons, as has been done in previous work on galaxies at closer
distances (e.g., Fall, Chandar, \& Whitmore 2005; Chandar et
al. 2010). We compare the predicted and measured H$\alpha$ maps for a
subsample of hand-selected young clusters and find that, unlike for
more nearby galaxies, no escape fraction is obviously needed.

In order to understand the impact that the escape fraction ($f_e$) has on the
cluster age estimation, we run the $UBVI$H$\alpha$ SED fitting with two values of
$f_e$; we use the best fit $f_e$ of 0.25, as well as $f_e=0.0$. We find
that $< 5\%$ of clusters have a difference in age of log$(\tau/\rm{yr}) >
0.1$ between age estimations from the two escape fractions, and adopt
$f_e=0.0$ for simplicity due to its negligible impact on age estimates.

\begin{figure}[ht]
\begin{center}
\includegraphics[scale=0.5]{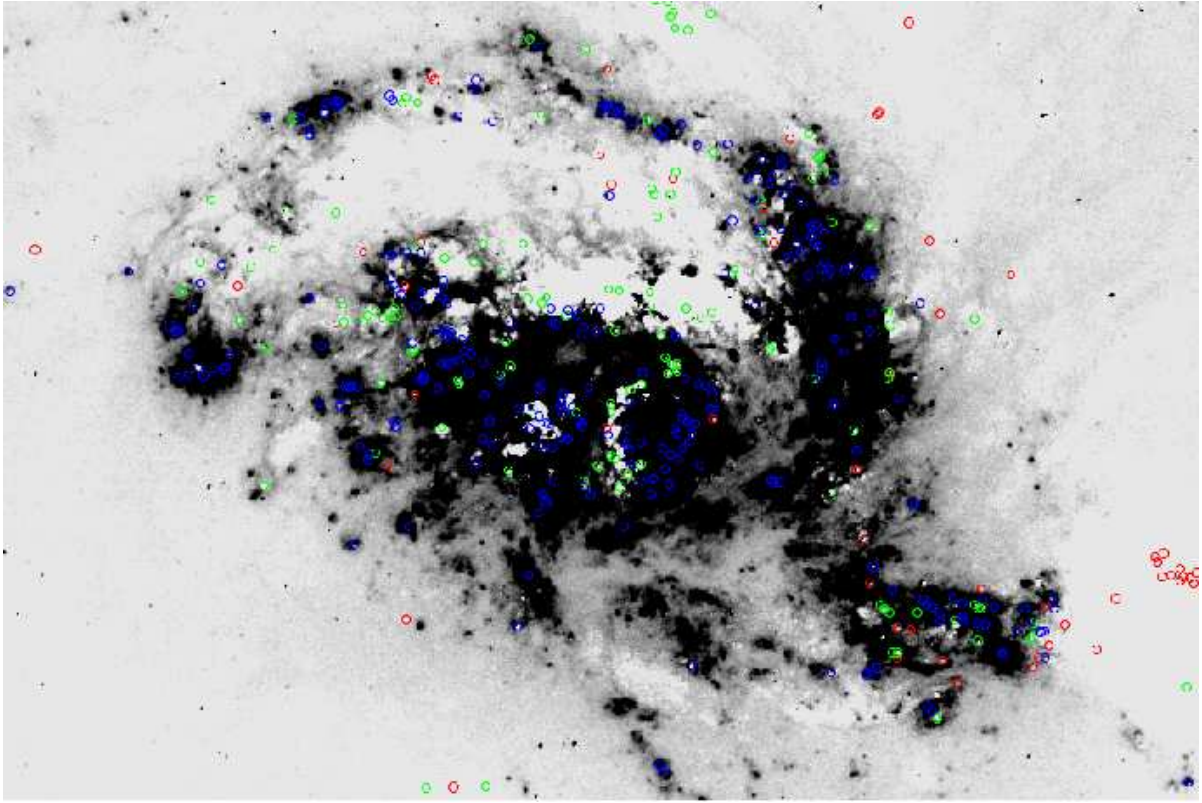}
\caption{A map of H$\alpha$ line emission, with continuum emission
  subtracted. The image has been inverted so that black regions
  show H$\alpha$ line emission. Blue, green, and red circles indicate locations of
  clusters with ages log$(\tau/\rm{yr}) < 7$, $7 <$ log$(\tau/\rm{yr})
  < 8$, and $8 <$ log$(\tau/\rm{yr}) < 8.6$, respectively.}
\label{ha_ages}
\end{center}
\end{figure}

We test that cluster ages are reasonable by performing some simple
checks. In most galaxies, it has been found that the brightest
clusters tend to be young (we discuss the physical implications of
this in Section \ref{CMF_SFR}; e.g., Whitmore et al. 1999). We test our
sample by picking the brightest 15 clusters after correction for
internal extinction and find that they are all younger than 10
Myr. In addition, Figure \ref{ha_ages} shows a continuum-subtracted
H$\alpha$ map of NGC 3256, with clusters overlaid that are color-coded
by age. Clusters younger than 10 Myr (blue circles) tend to clump
together and follow the brightest H$\alpha$ regions, while $10-100$
Myr old (green) and $100-400$ Myr old (red) old clusters are more
sparse and fall on areas with no H$\alpha$ emission.

\section{RESULTS}
\label{results}

\subsection{Luminosity Function}
\label{LF}

\begin{figure}[ht]
\begin{center}
\includegraphics[scale=1.3]{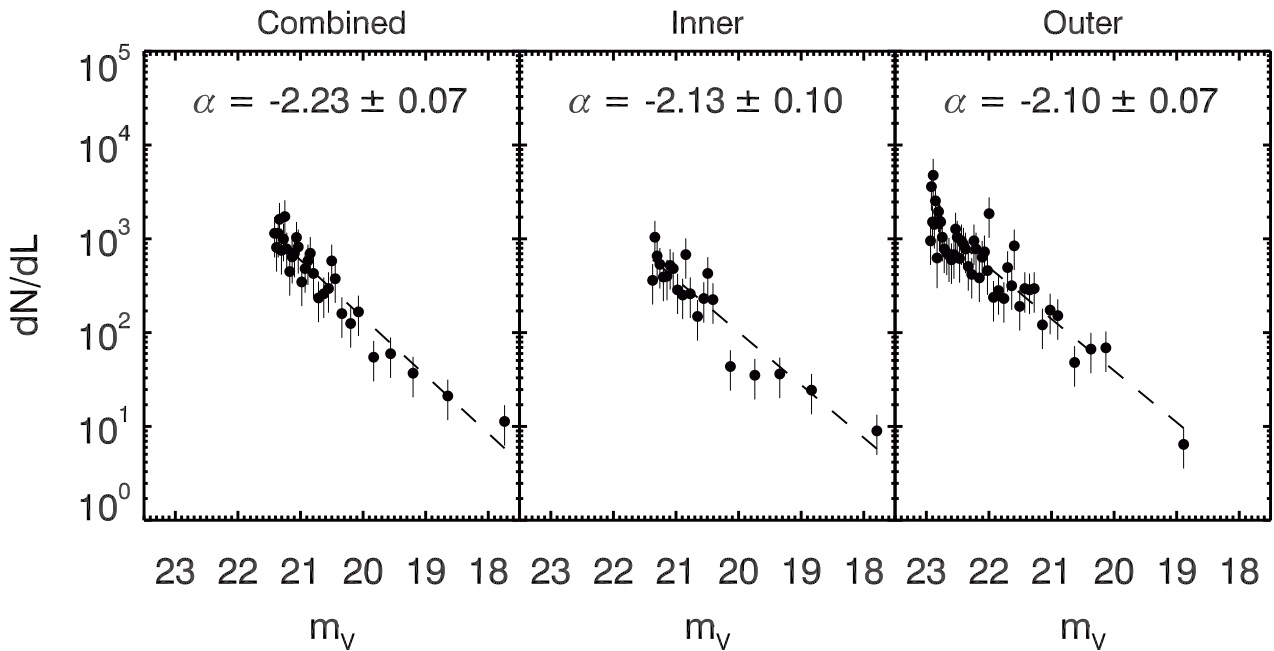}
\vspace{18.00mm}
\caption{Luminosity functions of NGC 3256, uncorrected for internal
  extinction within NGC 3256. On the left is the LF for all coverage
  of the main body. The middle panel shows the LF for the inner clusters,
  while the right panel shows the LF for the outer clusters.}
\label{dndl}
\end{center}
\end{figure}

We present the LFs for clusters in NGC 3256 in
Figure \ref{dndl}. The LF is best described by a power law, $dN/dL
\propto L^{\alpha}$, where $\alpha$ is determined from a linear fit to
log $(dN/d(m_V))$. The $m_V$ values have been aperture-corrected and
corrected for foreground extinction in the Galaxy. The three
panels of Figure \ref{dndl} show the LF for the entire sample, as well
as for clusters in the inner and outer regions of NGC 3256, separately. We find
$\alpha = -2.23 \pm 0.07$ for the combined sample with $m_V
\leq 21.5$. The LF for the inner and outer regions individually
yielded $\alpha \approx -2.1$ in both cases. We also determine the
power law index for the combined sample using $B$ and $I$ band
magnitudes, and find $\alpha = -2.14 \pm 0.09$ and $\alpha = -2.17 \pm
0.08$, respectively.  These are consistent with the $V$ band result of
$\alpha=-2.23$, within the uncertainties.

Zepf et al. (1999) previously presented a catalog of clusters in NGC
3256 using poorer quality (in both depth and resolution) $B$ and $I$
band Wide Field Planetary Camera 2 images from \emph{HST}. Using their
cluster catalog to measure the $B$ band LF, we find $\alpha = -2.15 \pm
0.07$, in agreement with the LF presented in this work. In addition,
our value for $\alpha$ is similar to that found in many other galaxies
(Whitmore et al. 2014).

We also compare the LF found in this work to that of clusters in the
eastern tidal tail of NGC 3256, as measured by Mulia, Chandar, \& Whitmore
(2015). They obtain $\alpha = -2.61 \pm 0.27$, somewhat steeper than the
LF in the main body, for clusters brighter than $M_V \approx -8.0$.

\subsection{Age Distribution}
\label{agedist}

\begin{figure}[ht]
\begin{center}
\includegraphics[scale=0.80]{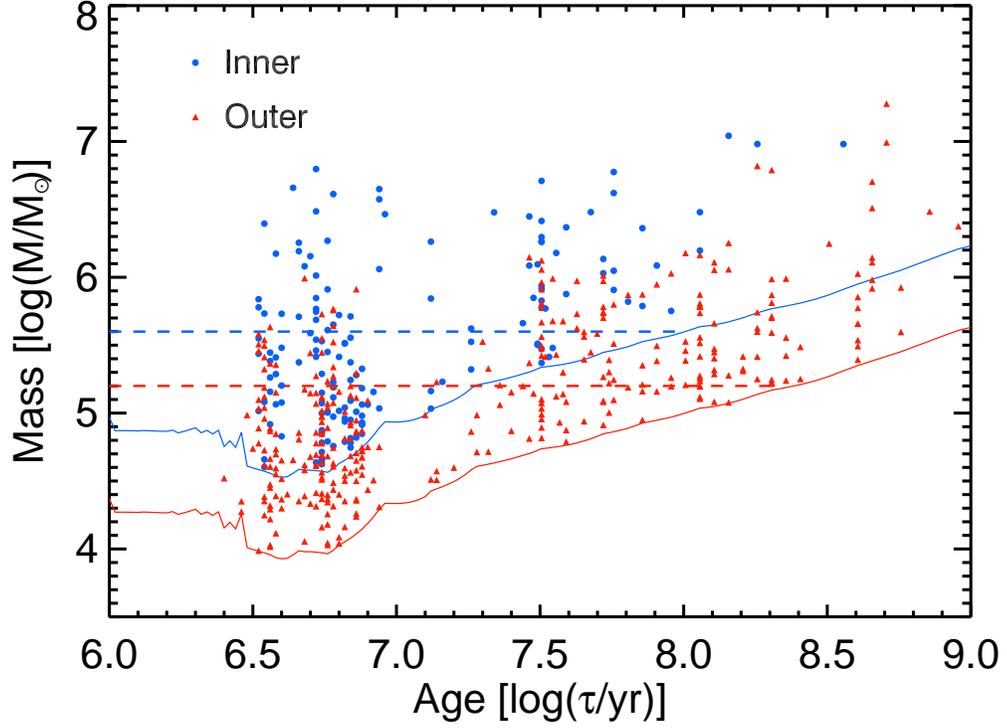}
\caption{Mass-age relation for clusters in the inner and
  outer regions of NGC 3256. The solid
  lines illustrate our completeness limits of $m_V=21.5$ (blue) and
  $m_V=23.0$ (red) for the inner and outer catalogs, respectively.}
\label{mass-age}
\end{center}
\end{figure}

\begin{figure}[ht]
\begin{center}
\includegraphics[scale=1.3]{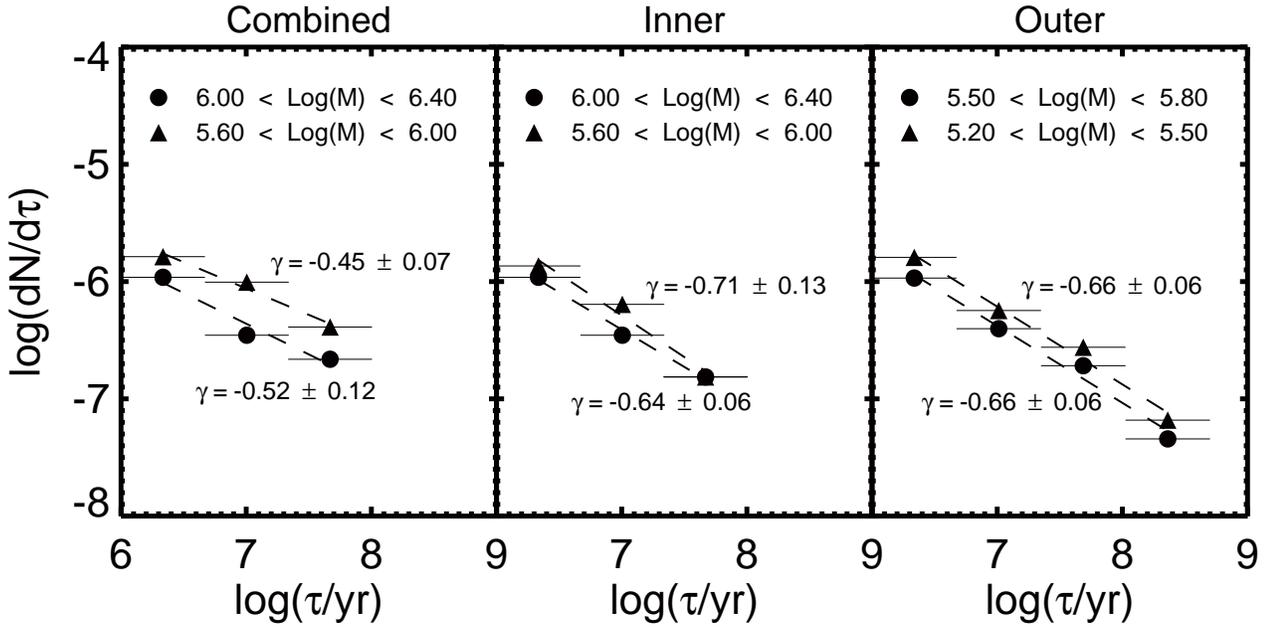}
\vspace{15.00mm}
\caption{Age distributions of NGC 3256 for two mass regimes, with best
  fits shown as dashed lines. Different panels are for different catalogs.}
\label{dndt}
\end{center}
\end{figure}

Figure \ref{mass-age} shows the mass-age relation, separated for
clusters in the inner and outer regions. The solid lines show our
brightness limits of $m_V = 23.0$ ($M_V = -9.79$ for a
distance modulus of 32.79) for the outer region and $m_V = 21.5$
($M_V = -11.29$) for the inner region. The different completeness
limits of the inner and outer regions result in different mass regimes
in which the age distribution can be determined. The two dashed lines
indicate the masses at which both samples are complete over the age
ranges used in the age distribution (Figure \ref{dndt}). The inner
cluster sample is complete to log($\tau/\rm{yr}) \approx 8.0$ for log($M/M_{\odot}$) $>
5.6$, while the outer sample is complete to log($\tau/\rm{yr}) \approx
8.4$ for log($M/M_{\odot}$) $> 5.2$. The apparent lack of clusters in the
range $7 <$ log$(\tau/\rm{yr}) < 7.5$ is an artifact of the SED fitting, partially
due to the BC06 track forming a loop in color space.

We find that the age distribution follows a power law with $dN/d\tau
\propto \tau^{\gamma}$, with $\gamma \approx -0.67 \pm 0.08$ for two
independent mass ranges in both the inner and outer regions. The combined
catalog from both regions yields $\gamma \approx -0.5$ for both mass
ranges; however, we note that the difference in completeness limits
for the two catalogs results in a very low number of clusters for which
the age distribution can properly be measured. The age
distribution presented in Figure 5 of G10 for clusters
in NGC 3256 shows a similar declining shape to that presented here.

Fall \& Chandar (2012) find the cluster age distribution for multiple
galaxies to be a power law with $\gamma$ between $\approx -0.6$ and
$-1.0$. While the slope of the age distribution fluctuates somewhat
depending on how the fit is performed, this fluctuation agrees well
with the fitting errors shown in Figure \ref{dndt}, as well as with the
uncertainties listed in Fall \& Chandar (2012). We conclude that
$\gamma$ for NGC 3256 is similar to the values obtained for the
cluster populations in other galaxies, although several works find
significantly flatter age distributions for clusters in some galaxies
(e.g., Hwang \& Lee 2010; Baumgardt et al. 2013; Silva-Villa et
al. 2014), including recent work on M31 by the PHAT team
  (Fouesneau et al. 2014).

The age distribution is a combination of the formation and
disruption histories of the clusters.  The star formation
histories (SFHs) are known independently in a number of galaxies
and appear to be relatively constant at least to within a factor of
two over the last few hundred million years (e.g., SMC -- Harris \& Zaritsky 2004;
LMC -- Harris \& Zaritsky 2009). Unfortunately, to our knowledge, there
is no published SFH for the merging, starbursting NGC 3256 system, either from
direct observations or from high quality simulations. If the SFH of NGC
3256 has been constant to within a factor of a few over this
timescale, as suggested by some simulations for the starbursting
Antennae galaxies (Karl, Fall, \& Naab 2011), as well as observational
works on interacting galaxies (e.g., Knapen, Cisternas, \& Querejeta
2015 and references therein), then the clusters in NGC 3256 would also
have a disruption history similar to that found in these more
quiescent galaxies.

\subsection{Mass Function}
\label{MF}

We show the CMFs in Figure \ref{dndm}. These can be
described by a power law, $dN/dM \propto M^{\beta}$.
We measure $\beta$ for three age ranges for clusters in the inner
and outer areas of the galaxy, as well as for the combined
sample. The given uncertainties only reflect the formal uncertainty
in the fit; we discuss other sources of error in Section \ref{MF_err}.

In the age range $6 <$ log$(\tau/\rm{yr}) < 7$, we find that the
combined sample of inner and outer clusters yields $\beta = -1.86 \pm
0.14$. Separating the sample yields $\beta = -1.84 \pm 0.14$
for outer clusters, but a shallower $\beta = -1.61 \pm 0.13$ for inner
clusters. The somewhat shallower slope is likely a result of
crowding, which can artificially flatten the mass function. Studies of other
galaxies have also found $\beta \approx -2$ (e.g., Zhang \& Fall 1999;
Chandar, Fall, \& Whitmore 2010; Chandar et al. 2010; Chandar, Fall,
\& Whitmore 2015). We note that G10 find $\beta = -1.85 \pm 0.12$
for the same age range in NGC 3256, quite similar to our result.

We only determine the mass functions for older clusters in
the outer region, since there are very few such clusters in the inner
region. We find $\beta = -1.31 \pm 0.36$ in the age range $7
<$ log$(\tau/\rm{yr}) < 8$. This age range is also the most difficult
to fit, partially due to the loop in the BC06 track discussed in
Section \ref{agedist}. Clusters with ages $8 <$ log$(\tau/\rm{yr}) <
8.6$ produce a mass function with $\beta = -2.08 \pm 0.45$, where the
large uncertainty in fitting is due to small-number statistics.

In Figure \ref{dndlogm_godd}, we also show the mass function
for NGC 3256 clusters younger than log($\tau/\rm{yr}) =7$ from the
G10 sample. We find that their mass function shows an unexpected systematic
offset toward lower masses. The offset is not due to differences in
photometry, in SSP model assumptions, or in derived age or extinction
values. When we put the G10 photometry through our age-dating
procedure, no offset exists. The reason for the offset between the G10
cluster mass estimates and our own is unclear.

\subsubsection{Uncertainties on $\beta$}
\label{MF_err}

Several factors contribute to the uncertainty in the
shape of the CMF. We mentioned some of these factors
already in Section \ref{ha_calib}, such as the impact that small
changes in the zero point and escape fraction for the H$\alpha$ filter
have on $\beta$. In this section, we consider additional contributors
to the uncertainties in $\beta$.

Incompleteness can impact $\beta$, causing it to become artificially
shallow. This is especially true when considering $\beta$ for a very
small age range, where brightness and mass are correlated, and a loss
of faint clusters is equivalent to a loss of low-mass clusters. We
discussed completeness of our catalog in Section \ref{clus_sel}, and we
choose selection limits that are sufficiently bright that the impact
of incompleteness is minimal.

Details of the fitting procedure can also affect $\beta$.
We find that the specific mass range used in the fits affects $\beta$
by less than about 0.05 for samples with 40 or more clusters, similar
to most of the subsets quoted in this paper. We also measure the impact of bin
size on $\beta$ and again find $\beta$ to fluctuate by $\pm
0.05$. Overall we find that all of these sources of uncertainty affect
$\beta$ within the formal uncertainty given in Figure \ref{dndm} for
each age range.

\subsubsection{Is There an Upper Mass Cutoff?}

Some works have found that mass functions for clusters in spiral
galaxies deviate from a single power law and have a downturn at the
high-mass end (e.g. Bastian et al. 2012). These works
suggest that the mass function is better described by a Schechter
function, $dN/dM \propto M^{\beta}$exp($-M/M_C$), where $M_C$ is the
upper mass cutoff. While the lower-mass end of the Schechter function
converges to a power law with index $\beta$, the
high-mass end drops off from a power law quickly above $M_C$. In NGC
3256, we find that a single power law provides a good fit to the
observed CMF, since no obvious downturn is observed
at the high-mass end. Therefore, our distribution is only consistent
with a Schechter function with a high $M_C$ (consistent with the
suggestion that any $M_C$ would be expected to be higher than $10^6$
$M_{\odot}$ for this type of system; e.g., Kruijssen 2014). Below, we
determine a lower limit to $M_C$. For simplicity, we compare our CMFs for $6 <$
log$(\tau/\rm{yr}) < 7$ and $8 <$ log$(\tau/\rm{yr}) < 8.6$ to
Schechter functions with a fixed $\beta = -2.0$ and
varying $M_C$ values, although there is no reason why $\beta$ has to be
exactly $-2.0$. For the latter age range, we use only the sample taken
from the outer region of the central galaxy due to low-number statistics.

Using this approach, the mass function of clusters with ages $6 <$
log$(\tau/\rm{yr}) < 7$ can only be reasonably fit for a Schechter
function with log$(M_C/M_{\odot})$ values above 6.5. Because the mass
function is fit out to log$(M/M_{\odot}) \sim 5.75$, it is not
surprising that Schechter functions with log$(M_C/M_{\odot}) < 6.5$
begin to fall below the observed mass function at the high-mass end. The left
panel of Figure \ref{dndm_schec} shows the mass function with multiple
$M_C$ values.

Fitting the mass function for $8 <$ log$(\tau/\rm{yr}) < 8.6$ with a Schechter
function is indistinguishable from fitting a pure power law for
log$(M_C/M_{\odot}) \geq 6.5$. This is shown in the right panel of
Figure \ref{dndm_schec}, where log$(M_C/M_{\odot}) = 6.0$ and
log$(M_C/M_{\odot}) = 6.5$ provide reasonable fits to the mass function,
as does a single power law. The mass function is fit out to
log$(M/M_{\odot}) \sim 5.5$, and we find
that the Schechter functions with log$(M_C/M_{\odot}) < 6.0$ begin to
fall below the observed mass function at the high end.

We find little evidence for a truncation of the mass function at the
high end. We conclude that while both mass functions tested here can
be fit with a Schechter function, it is not required and that a single
power law provides an equally good fit to each. We therefore favor a
pure power law in fitting the mass functions.

\section{How Does Distance Affect the Observed Cluster Distributions?}
\label{dist}
The specific method used to select star clusters, particularly at ages
$\tau < 10$ Myr, may affect the results (e.g., Bastian et
al. 2012). In addition, cluster selection in galaxies located at different
distances will necessarily be limited by resolution since at least
some sources will appear as diffuse stellar clusters in nearby
galaxies but not in more distant ones.

At a distance of 36 Mpc, NGC 3256 is nearly twice as far as the
Antennae, the nearest example of a pair of actively merging
galaxies. In order to assess how distance might impact source
selection, as well as the LFs, mass functions, and age distributions
of cluster populations in NGC 3256, we use the well studied \emph{HST}
images of the Antennae, and compare the resulting cluster
distributions when we simulate the entire system to lie a factor of
two further away.

We download the Antennae data from \emph{HST}/WFC $F435W$, $F550M$,
and $F814W$ filters, as well as the WFC 3 $F336W$ filter
from the HLA. The combined image for each filter is then boxcar-smoothed with a
kernel width of two pixels. We run DAOFIND on both the original and
degraded $F550M$ images, tuning only the FWHM parameter in both images,
and we detect 8405 and 3173 objects, respectively. We refer to the cluster
catalogs produced from the unaltered and degraded images as the
``original'' and ``image-smoothed'' catalogs, respectively. We
perform aperture photometry on all detected sources and correct for foreground
extinction. In addition, we measure and apply separate aperture
corrections to each catalog. 

Figure \ref{ant_image} shows a comparison of the original and
image-smoothed $F550M$ images. The circles indicate detections from
DAOFIND. It is evident that the image-smoothed
source catalog is missing many of the very faint clusters in addition
to some of the bright sources. We illustrate this in Figure
\ref{ant_dndl}, where the bottom panel shows the relative fraction of
sources found in the image-smoothed catalog compared to the original
catalog.\footnote{We do not restrict our sample to objects that
  appear to be spherically symmetric, as has been recommended to
  separate bound from unbound clusters (e.g., Gieles \& Portegies
  Zwart 2011), since resolution makes this difficult at the distance
  of the Antennae and NGC 3256.} We find that $\approx 85\%$ of the 3173
image-smoothed detections share coordinates (within one pixel) with their original
image counterparts.

The top and middle panels of Figure \ref{ant_dndl} show the LFs for
the original and image-smoothed cluster catalogs, respectively. Both LFs are fit for
$m_V < 22.7$ ($M_V < -9$), where the $m_V$ distribution begins to
flatten out at the faint end. We find values for $\alpha$
of $-2.13 \pm 0.04$ for the clusters detected in the original image, very
similar to those found in previous works. For the smoothed image
sample we also find $\alpha = -2.10 \pm 0.04$. A similar and more
extensive study was performed by Randriamanakoto
et al. (2013), where it was also determined that $\alpha$ is not
greatly impacted by resolution. They performed various convolutions on
their image in order to simulate different galaxy distances, finding
that $\alpha$ differed by no more than 0.2.

The color distributions of clusters brighter than $M_V < -9$ from the
original and smoothed images are quite similar. While there are more
points from the original catalog present in Figure \ref{ant_CCD}, the
scatter in the colors among the two catalogs is roughly the
same. We run SED fitting on both catalogs and construct their age
distributions (not shown), finding $\gamma \approx -0.8$ in both cases.

We find that the LF, color distributions, and age distributions for
the original and image-smoothed catalogs are nearly the same, and we
therefore conclude that distance does not significantly hamper our
ability to study clusters out to $\approx 40$ Mpc. 

\section{How Efficiently Does NGC 3256 Form Clusters?}
\label{discussion}

Many previous works have suggested that galaxies with a high
SFR per unit area ($\Sigma_{\rm{SFR}}$) form clusters
more efficiently than galaxies with lower $\Sigma_{\rm{SFR}}$ (e.g.,
G10; Kruijssen 2012; Kruijssen \& Bastian
2016). Two independent observational methods have
been developed to test this: the CMF/SFR statistic (Chandar, Fall, \&
Whitmore 2015; hereafter CFW15) and $\Gamma$, the fraction of 
stars formed in bound clusters (Bastian et al. 2008). A third, less
direct method, $T_{L}(U)$, has also been used (Larsen \& Richtler
2000). In this work, we focus on the first two methods, because they are
easier to interpret. In this section,
we apply both methods to our cluster sample in NGC 3256,
and compare the results with those from previous works and for
different galaxies.

\subsection{CMF/SFR Statistic}
\label{CMF_SFR}

CFW15 compared the CMFs between
several different galaxies by normalizing them by the SFR of the host
galaxy, and comparing the amplitudes of the resulting CMF/SFR statistic. The
sample included spirals (M83 and M51), irregulars and starbursts
 (Large and Small Magellanic Clouds, NGC 4214, and NGC 4449), and
the merging Antennae galaxies.  The large range of amplitudes for the
CMFs in two different ranges of age (log$(\tau/\rm{yr}) < 7$ and $8 <$
log$(\tau/\rm{yr}) < 8.6$) in these galaxies collapse when divided
by their respective SFRs. The dispersion in the
CMF/SFR statistic is only $\sigma \approx 0.2$, similar to the
expected uncertainties. Since the CMF/SFR statistic of very young 
clusters is similar across these different galaxies, 
CFW15 concluded that cluster formation scales with the overall SFR on
galaxy scales. The similarity of the CMF/SFR statistic for the older clusters
suggests that the disruption histories are also similar across the
galaxies.

Here, we determine the CMF/SFR statistic for NGC 3256 for the
first time and compare with the rest of the galaxies in the CFW15
sample. The mass functions discussed in Section \ref{MF} were shown
to follow a power law with index $\beta$. In the left panel of Figure
\ref{dndlogm}, we show the unnormalized CMFs for the same seven galaxies
shown in CFW15 and now also include NGC 3256. The CMF for NGC 3256 is
higher than that of the other galaxies, including the Antennae. 
The right panel of Figure \ref{dndlogm} presents the CMF/SFR statistic
for NGC 3256 and shows that it is similar to that of the other seven
galaxies. We measure the CMF/SFR statistic only for clusters younger than
log$(\tau/\rm{yr}) < 7$, because our NGC 3256 sample does
not contain a significant number of clusters older than this.

The uncertainties in the CMF/SFR statistic can be separated into
uncertainties in the CMF and in the SFR value. Here we focus on the
latter and refer the reader to Section \ref{MF_err}, where we
discussed uncertainties regarding the mass function. 
We adopt a total SFR of 50 $M_{\odot}$ yr$^{-1}$ for NGC 3256 (Sakamoto et
al. 2014), which was calculated from the bolometric infrared
luminosity. This SFR is quite similar to the value of 46
$M_{\odot}$ yr$^{-1}$ calculated from the total infrared luminosity and
adopted in G10 for the SFR. We find other modern calculations of the
SFR in NGC 3256 in a similar range, with a high
value of 85 $M_{\odot}$ yr$^{-1}$ using infrared luminosity
(Rodr{\'i}guez-Zaur{\'i}n et al. 2011). These values are consistent
with an accuracy to within a factor of about two for the SFR, as
found by several previous works (e.g., Lee et al. 2009; CFW15).

While our full area of coverage is $\approx 120$ kpc$^2$, the clusters
fall within an area of $\approx 50$ kpc$^{2}$,
comparable to that used for the SFR determinations mentioned above.
While it is important that the same area of the galaxy be used for
determining both the CMF and the SFR (as done in CFW15), this has not
always been done.  A number of studies have compared cluster
properties with the SFR density of
the host galaxy, but with the latter determined for a different area
than for the clusters.  For NGC 3256, G10 assumed a star-forming area
of $\approx 75$ kpc$^2$ (taken from the literature), but the actual
area covered by their cluster catalog is only $\sim33$ kpc$^2$ (see
Section \ref{clus_sel} and Figure \ref{image}). We estimate the
uncertainty in previously published values of the star-forming
\emph{area} to be approximately a factor of two for NGC 3256. This is
on top of the uncertainty in the SFR itself.

CFW15 quantify the scatter in the CMF/SFR statistic among galaxies 
by fitting each galaxy by

\begin{equation}
dN/dM = A \times \rm{SFR} \times (M/10^4 M_{\odot})^{-1.9}
\label{CMF_SFR_eq}
\end{equation}

\noindent and find the scatter in the normalization coefficient $A$ to be
$\sigma($log $A) = 0.21$. When NGC 3256 is added as an eighth galaxy, we
find $\sigma($log $A) = 0.24$.

If the formation of clusters is affected by the overall SFR or
$\Sigma_{\rm{SFR}}$ of their host galaxy, we should see a
correlation of these properties with the residuals of log $A$. In
Figure \ref{residuals}, we plot the residuals of log $A$
for each galaxy versus the SFR and $\Sigma_{\rm{SFR}}$. We take
$\Sigma_{\rm{SFR}}$ values for all galaxies (including NGC 3256), but
not the Antennae, from the recent compilation made by Adamo et
al. (2015). The result for NGC 3256 is shown in red. The
left panel is similar to the upper-left panel of Figure 4 in CFW15,
who did not show the residuals with respect to $\Sigma_{\rm{SFR}}$.
No obvious trend in the residuals is observed with either the SFR or
with $\Sigma_{\rm{SFR}}$. The nonparametric Spearman correlation
coefficient, $r_{\rm{S}}$, and the corresponding probability $p(>r_{\rm{S}})$
confirm this visual impression. We find $r_{\rm{S}}$ of -0.07 and
$p(>r_{\rm{S}})$ of 0.87 for log $A$ versus SFR and $r_{\rm{S}}$ of -0.24 and
$p(>r_{\rm{S}})$ of 0.57 for log $A$ versus $\Sigma_{\rm{SFR}}$. These
$p(>r_{\rm{S}})$ values should be less than 0.05 to have a greater
than 95\% confidence of there being a correlation. Therefore, there is no
statistically significant correlation between the residuals in log $A$
and SFR and $\Sigma_{\rm{SFR}}$ for these eight galaxies.

Our results for NGC 3256, when added to those for the seven galaxies
presented in CFW15, are consistent with the hypothesis that the
formation rates of stars and clusters are similar across star-forming
galaxies that span a large range (approximately three orders of
magnitude) in SFR and $\Sigma_{\rm{SFR}}$. Our results for the
age distribution in NGC 3256, which also shows a declining shape
similar to that in the other more quiescent galaxies, suggest that
the disruption history may also be the same, although this conclusion
depends on the (currently unknown) SFH of the system. The cluster
population models presented in Whitmore, Chandar, \& Fall (2007)
indicate that for cluster disruption that is mostly independent of
mass, the youngest clusters are the brightest. This is consistent with
our findings for NGC 3256 in Section \ref{ha_calib}.

\subsection{The $\Gamma$ Statistic}
\label{Gamma}

A second, similar statistic that has been widely used in the
literature is $\Gamma$, defined as the mass formed in bound clusters divided
by the total stellar mass. The total mass in clusters, within a given age
range, is typically found by extrapolating the observed CMF down to a
mass of 100 $M_{\odot}$. This extrapolation is
accomplished either by assuming a power law or Schechter function, or
by using a simulated cluster population. $\Gamma$ has been
estimated for a number of galaxies and is typically plotted against
$\Sigma_{\mathrm{SFR}}$ of the host galaxy. Previously, both
observational and theoretical works have suggested that there is a
strong correlation between $\Gamma$ and $\Sigma_{\rm{SFR}}$ (and
$T_{L}(U)$ and $\Sigma_{\rm{SFR}}$), such that
the higher the $\Sigma_{\rm{SFR}}$ of a galaxy, the higher the
fraction of stars formed in bound clusters. Since estimates of
$\Sigma_{\rm{SFR}}$ and SFR correlate strongly for the specific galaxies studied
thus far, there should be a similarly strong correlation between
$\Gamma$ and SFR. If confirmed, this result would differ from the
result of CFW15 who found little variation from galaxy to galaxy for CMF/SFR.

G10 estimated $\Gamma$ for NGC 3256, and found $\Gamma = 22.9\%
^{+7.3}_{-9.8}$. This was calculated from the CFR divided by the SFR,
and it included only clusters with log($\tau/\rm{yr}) < 7$. We
summarize their method below, and refer the reader to G10
for details. G10 found the total mass in clusters above a cutoff of
log$(M_{\rm{cut}}/M_{\odot})=4.7$ and used a synthetic cluster
population to find the fraction of mass contained in clusters that is
below this cutoff, assuming a power law mass function with slope
$\beta=-2.0$. They obtained the CFR by dividing their total mass in
clusters by 7 Myr, rather than 10 Myr, due to the expectation that the
clusters are embedded for the first 3 Myr and hence are missing from
the sample. They divided the CFR by the SFR and obtained $\Gamma =
0.12$. They made three additional corrections, each increasing
$\Gamma$ by $\sim25$\%, giving a final value of $\Gamma=0.23$. G10
made these corrections in order to account for: (1) their rejection of
clusters with poor fits and photometry, (2) a range of metallicities
for young clusters in NGC 3256, which G10 found would increase the
total cluster mass relative to their assumptions, and (3) an
underestimate in actual cluster masses relative to derived masses.

We follow the G10 method, for the most part, to estimate
$\Gamma$ from our larger cluster catalog in NGC 3256.
We find the total mass in clusters above
log$(M_{\rm{cut}}/M_{\odot})=5.0$ and younger than 10 Myr to be $7.23
\times 10^7 M_{\odot}$. We use a similar synthetic cluster population as G10 to calculate the
fraction of mass contained in clusters for $M(M>M_{\rm{cut}})$. We
arrive at a CFR of 16.7 $M_{\odot}$ yr$^{-1}$ by
dividing $M_{\rm{tot}}$ by 7 Myr, and we calculate $\Gamma =$ CFR/SFR
$= 0.33$, when we assume an SFR of 50 $M_{\odot}$ yr$^{-1}$. We do not
make the three 25\% corrections in our determination of $\Gamma$,
because our procedure does not result in the rejection of clusters
with poor fits, we do not find any direct information on the
metallicity distribution of young clusters in NGC 3256, and we are
using different SSP models that, as far as we know, do not
systematically underestimate cluster masses. Our $\Gamma$ value of
0.33 is $2.75$ times higher than the value of 0.12 obtained by G10
before they make the additional corrections described above. This
factor is the same as the relative offset in the masses that we found
in Section \ref{MF}. Our results suggest that $\Gamma$ is fairly
robust when the same set of assumptions are used.

Ideally, $\Gamma$ should be determined using the same set of
assumptions for all galaxies. However, this is not the current state
of $\Gamma$ estimates in the literature, where different sets of
assumptions have been used for different galaxies. For completeness,
in Table \ref{gammas}, we compile different estimates of $\Gamma$ for NGC
3256 using different sets of assumptions. Here, column 4 shows our
`best' estimates of $\Gamma$, while column 5 includes the three 25\%
corrections used by G10 (labeled as $\Gamma'$). We find that one of the
biggest sources of uncertainty is the exact value used for the power
law index of the CMF, $\beta$. G10 and other works
have found that $\beta$ has typical uncertainties of the order of $\pm 0.2$.
However, changes at this level have a strong impact and change the
derived value of $\Gamma$ by a factor of nearly four. $\Gamma$ is also
sensitive to both the lower and upper mass ranges assumed in the
extrapolation. If $\beta=-2.0$, then each decade in mass adds an equal
percentage of total mass. For example, extrapolating down to a lower
mass cutoff of, say, $1000 M_{\odot}$ rather than $100 M_{\odot}$
would increase the fraction of $M(M>M_{\rm{cut}})$ by $\sim 10\%$. 
The estimates of $\Gamma$ vary from 0.09 to 0.98 (i.e., a factor of approximately
10!), depending on the specific choice of parameters. If we apply
the same three 25\% corrections as used by G10, we find a range
between 0.18 and 1.91, i.e. the estimates go higher than 100\%.

As also found by G10, the exact value of $\Gamma$ is quite
sensitive to the specific assumptions and
extrapolations that are made. Because a variety of assumptions have
been made for determinations of $\Gamma$ in different galaxies (which
we will address in a forthcoming paper), and because this method
requires more assumptions than the CMF/SFR statistic, the current
determinations have larger uncertainties, which may make it more
difficult to compare results between galaxies.
 
\section{SUMMARY AND CONCLUSIONS}
\label{conc}

We have used ACS/WFC from \emph{HST} to measure the properties of star
clusters in the main body of NGC 3256. We draw the following conclusions.

\begin{figure}[ht]
\begin{center}
\includegraphics[scale=1.3]{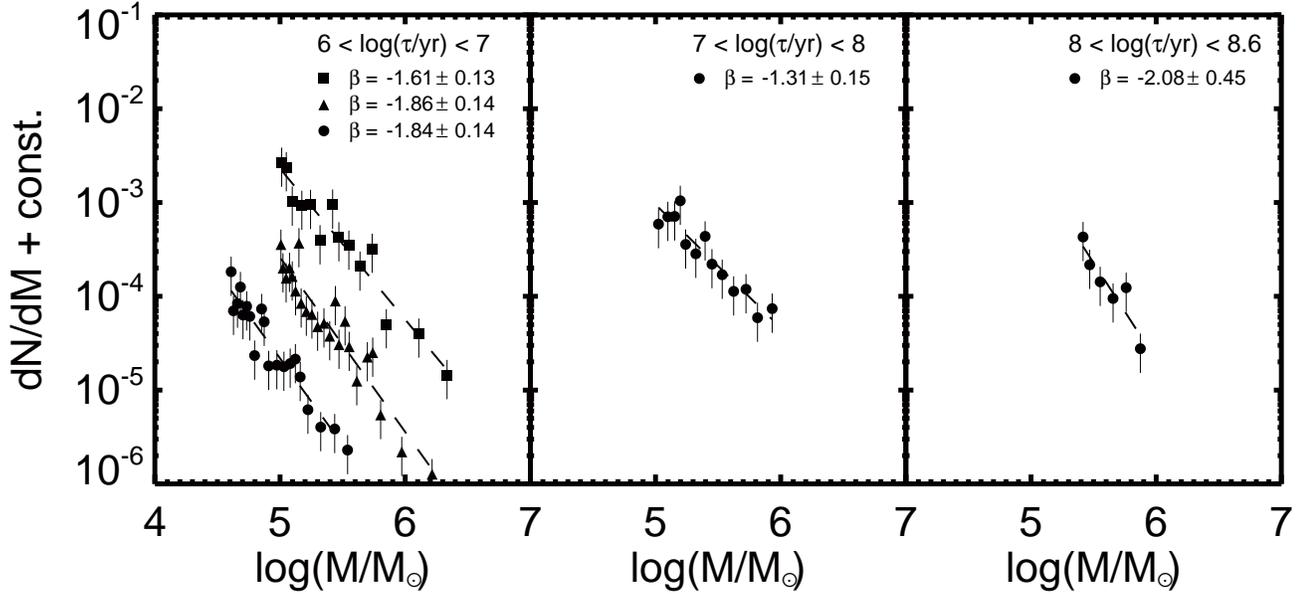}\\
\vspace{20.00mm}
\caption{Cluster mass functions for NGC 3256 for three age
  ranges. Different symbols indicate catalogs from the inner area of
  the main body (squares),
  the outer area (circles), and both catalogs combined
  (triangles). The inner and combined catalogs contained too few
  clusters for reasonable fits to mass functions for ages
  log$(\tau/\rm{yr}) > 7$ and are excluded from the middle and right
  panels. Power law fits to the mass functions are done for a constant
  number of clusters per bin. We find that a power law provides a good
  fit to the cluster mass functions in all cases. See Section \ref{MF}
  for details.}
\label{dndm}
\end{center}
\end{figure}

\begin{figure}[ht]
\begin{center}
\includegraphics[scale=0.7]{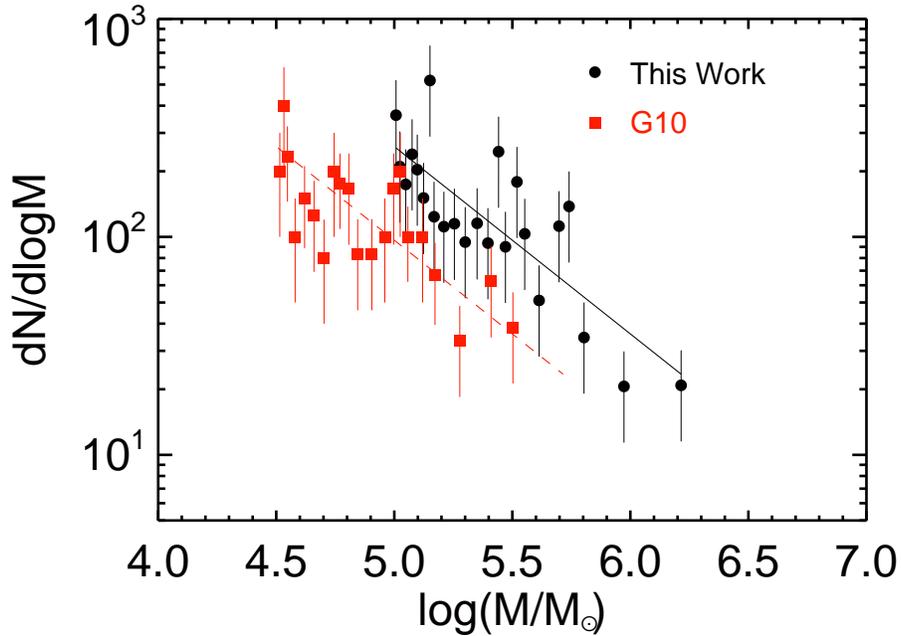}\\
\vspace{3.00mm}
\caption{Cluster mass functions for log$(\tau/\rm{yr}) <
  7$ clusters in NGC 3256. The mass function constructed from the
  catalog from G10 is shown as red squares and compared to the mass function
  used in this work (shown in black). The black line shows the fit to our mass
  function. The dashed red line, which serves only as a visual guide to
  illustrate the horizontal offset, is the same as the solid black line but offset
  leftward by 0.5, showing that the two mass functions have similar slopes.}
\label{dndlogm_godd}
\end{center}
\end{figure}

\begin{figure}[ht]
\begin{center}
\includegraphics[scale=1.1]{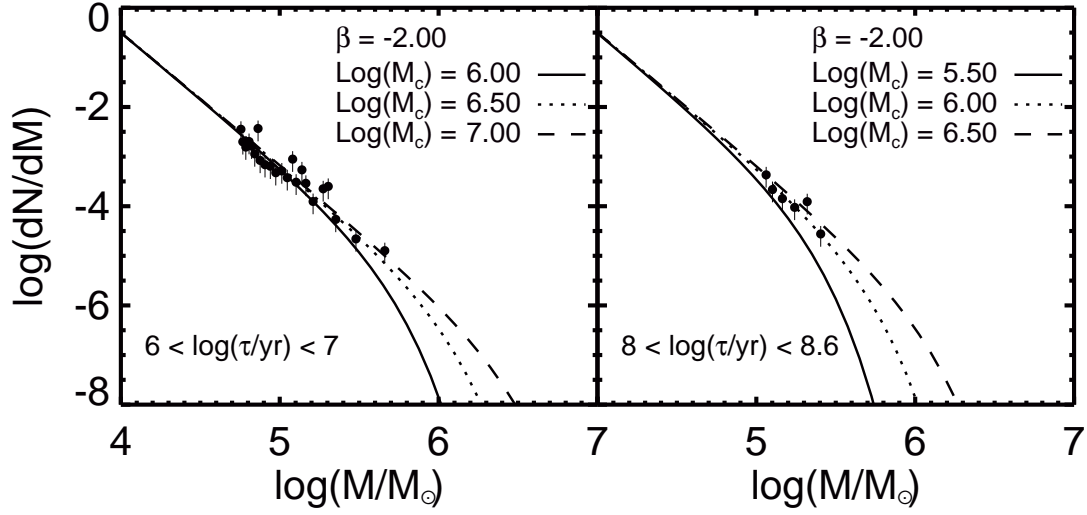}
\vspace{15.00mm}
\caption{Cluster mass functions fit with Schechter functions,
  represented by $dN/dM \propto M^{\beta}$exp($-M/M_C$). On the left,
  clusters in the range $6 <$ log$(\tau/\rm{yr}) < 7$ are well fit by
  $\beta = -2$ and log$(M_C) \gtrsim 6.5$. On the right, the mass
  function for $8 <$ log$(\tau/\rm{yr}) < 8.6$ is fit by $\beta=-2$ and
  log$(M_C) \gtrsim 6.0$. We find that the Schechter function is not
  necessary to fit the mass function (see Figure \ref{dndm}).}
\label{dndm_schec}
\end{center}
\end{figure}

\begin{figure}[ht]
\centering$
\begin{array}{cc}
\includegraphics[scale=0.22]{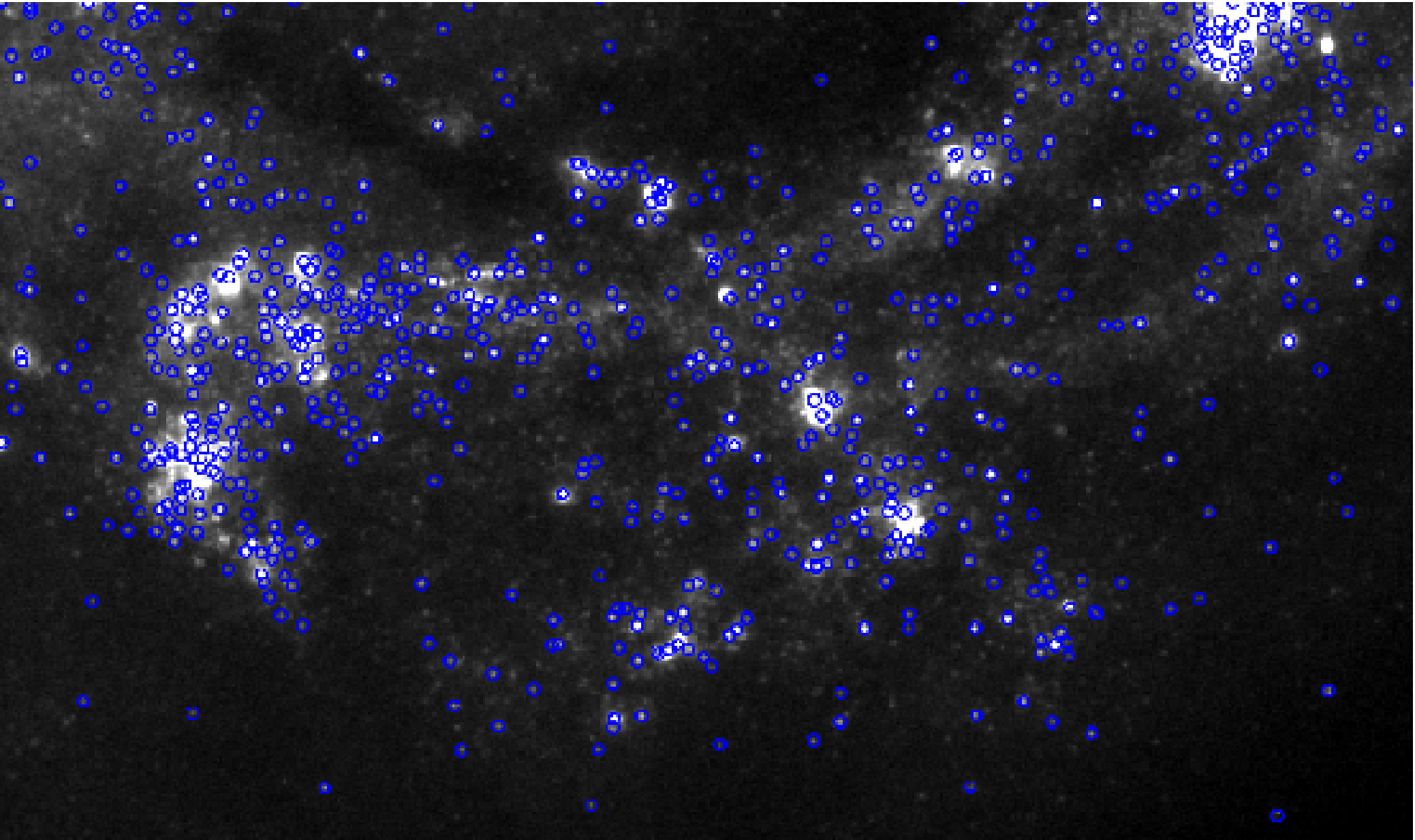}&
\includegraphics[scale=0.22]{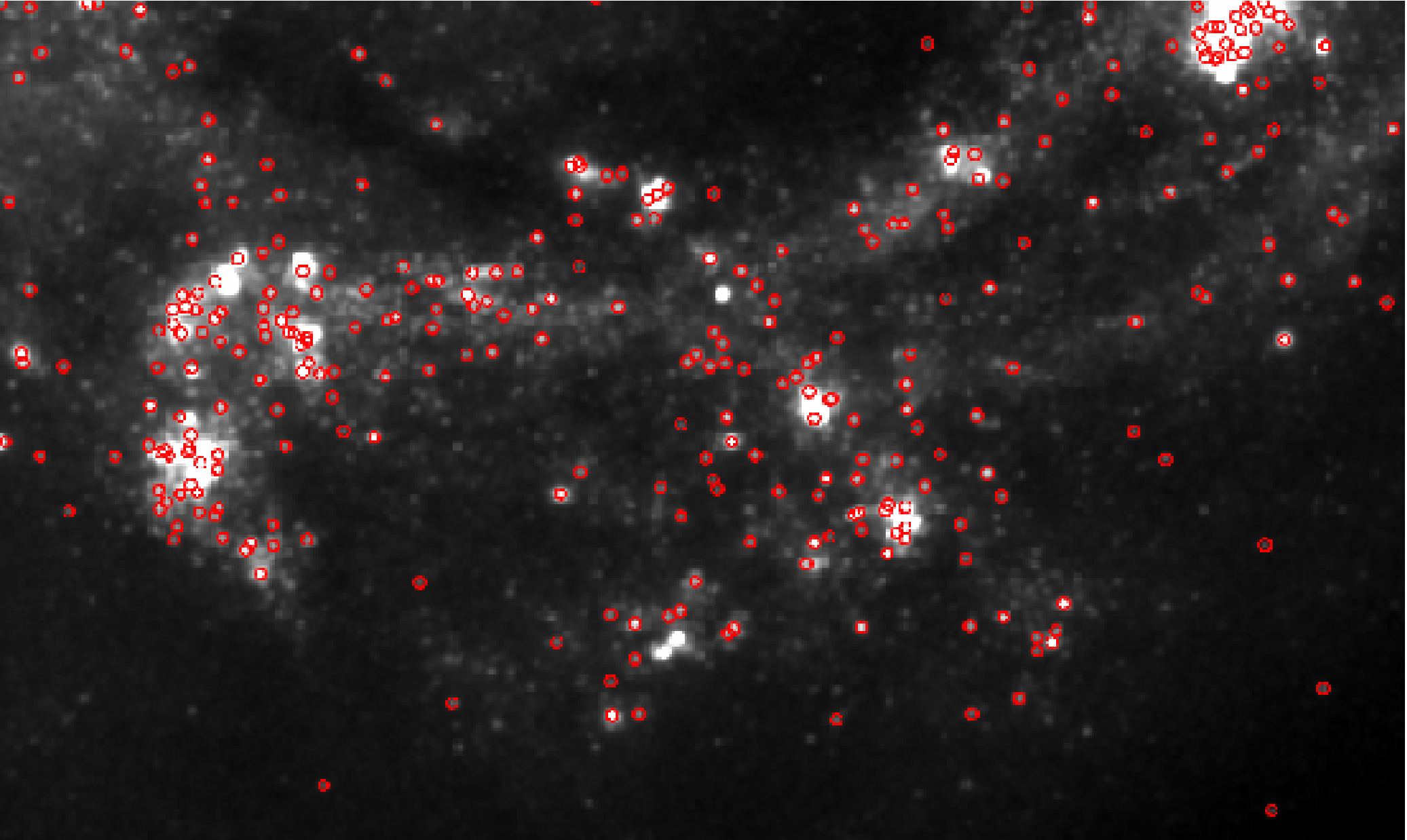}\\
\end{array}$
\caption{$F550M$ images of a portion of the Antennae taken with
  \emph{HST}/WFC at the native resolution ($0.05''$/pixel, left) and
  smoothed by two pixels to simulate the distance of NGC 3256
  (right). Circles indicate source detections from DAOFIND. By
  smoothing the image, many faint sources and sources in crowded areas
  are not detected. Despite this, the resulting LF and color
  distributions are quite similar.}
\label{ant_image}
\end{figure}

\begin{figure}[ht]
\begin{center}
\includegraphics[scale=1.0]{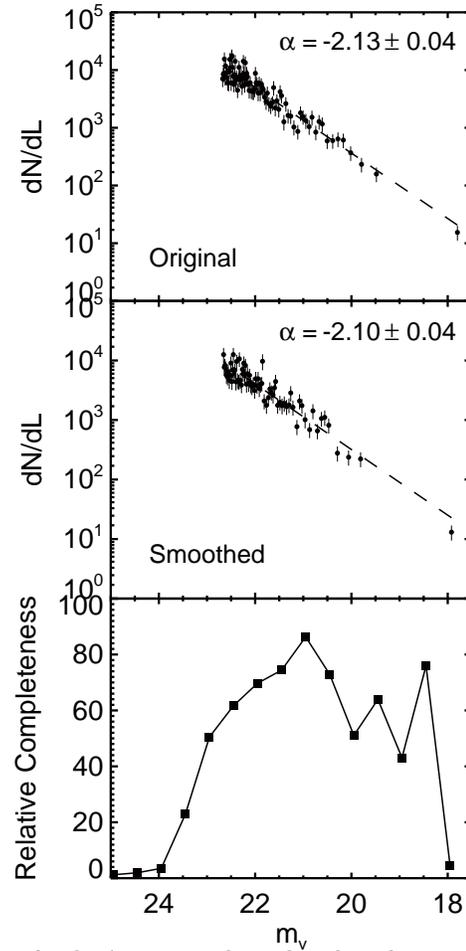}
\vspace{5mm}
\caption{Top: the luminosity function for the Antennae
  galaxies based on clusters detected in the $V$ band image of
  \emph{HST}/ACS. Middle: same as the top panel, but the LF
  resulting from detections made in the smoothed image. Note that
  nearly the same $\alpha$ value is obtained in both
  cases. Bottom: relative fraction of sources found in the
  image-smoothed catalog compared to the original catalog.}
\label{ant_dndl}
\end{center}
\end{figure}

\begin{figure}[ht]
\begin{center}
\includegraphics[scale=1.0]{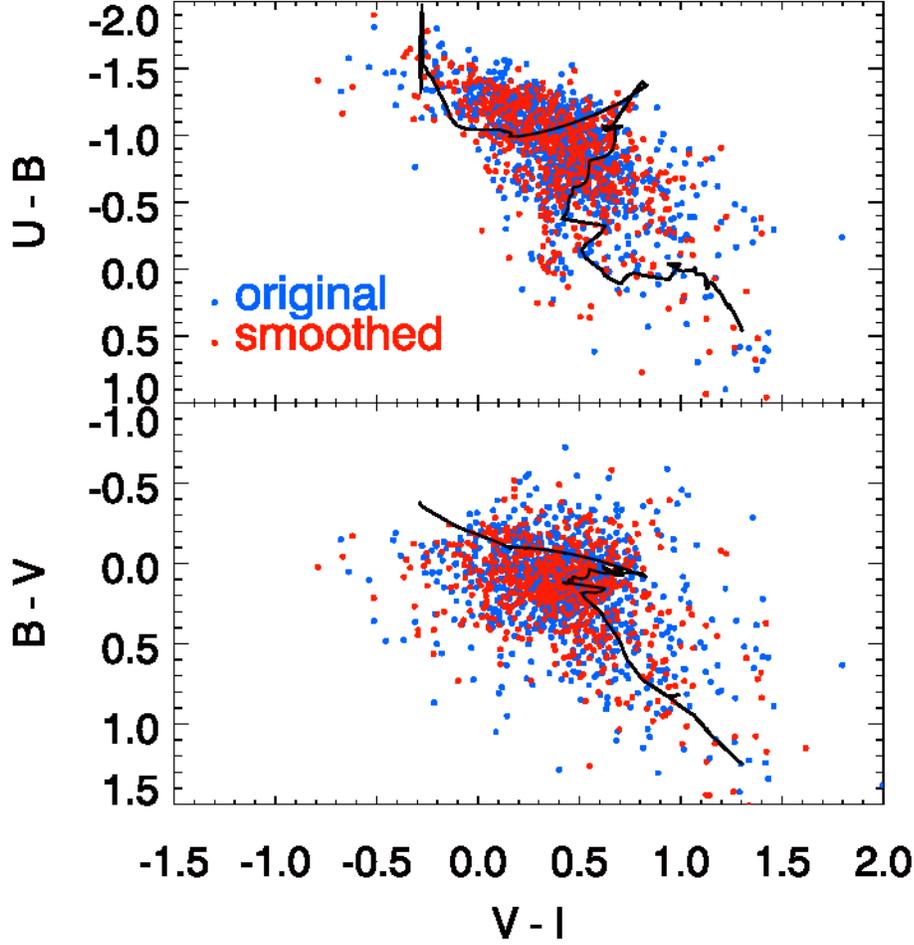}
\caption{Color-color diagrams of the Antennae for clusters with $M_V <
  -9$. Blue and red points indicate colors from the original and
  image-smoothed catalogs, respectively.}
\label{ant_CCD}
\end{center}
\end{figure}

\begin{figure}[ht]
\begin{center}
\includegraphics[scale=0.7]{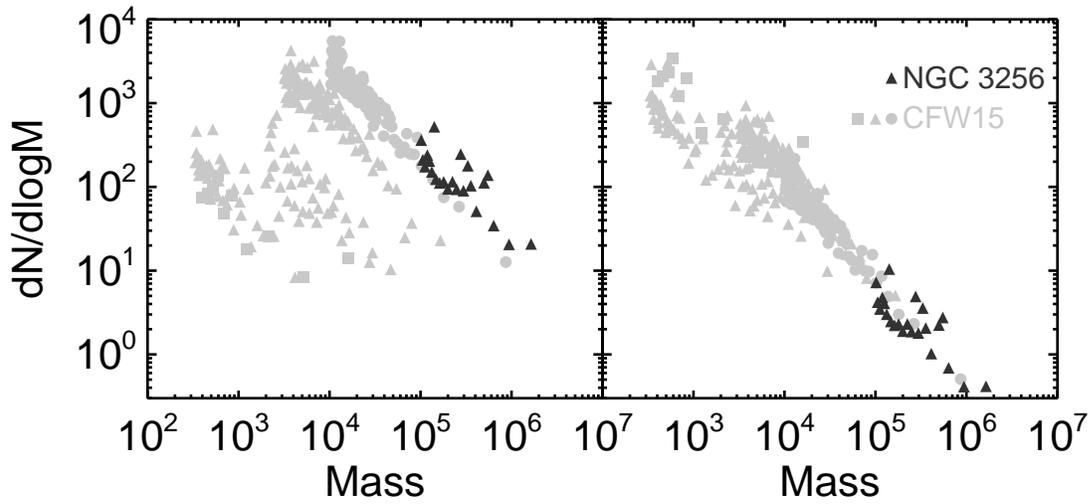}
\vspace{15mm}
\caption{Left: the unnormalized mass function for NGC 3256
  clusters younger than log$(\tau/\rm{yr}) < 7$, shown in dark gray, while the
  lighter gray points show the unnormalized mass functions of seven
  other galaxies calculated in CFW15. Right: same as the left
  panel, with mass functions divided by the SFR of their host
  galaxy. NGC 3256 falls on top of and extends the CMF/SFR
  distributions in the other seven galaxies.}
\label{dndlogm}
\end{center}
\end{figure}

\begin{figure}[ht]
\begin{center}
\includegraphics[scale=0.7]{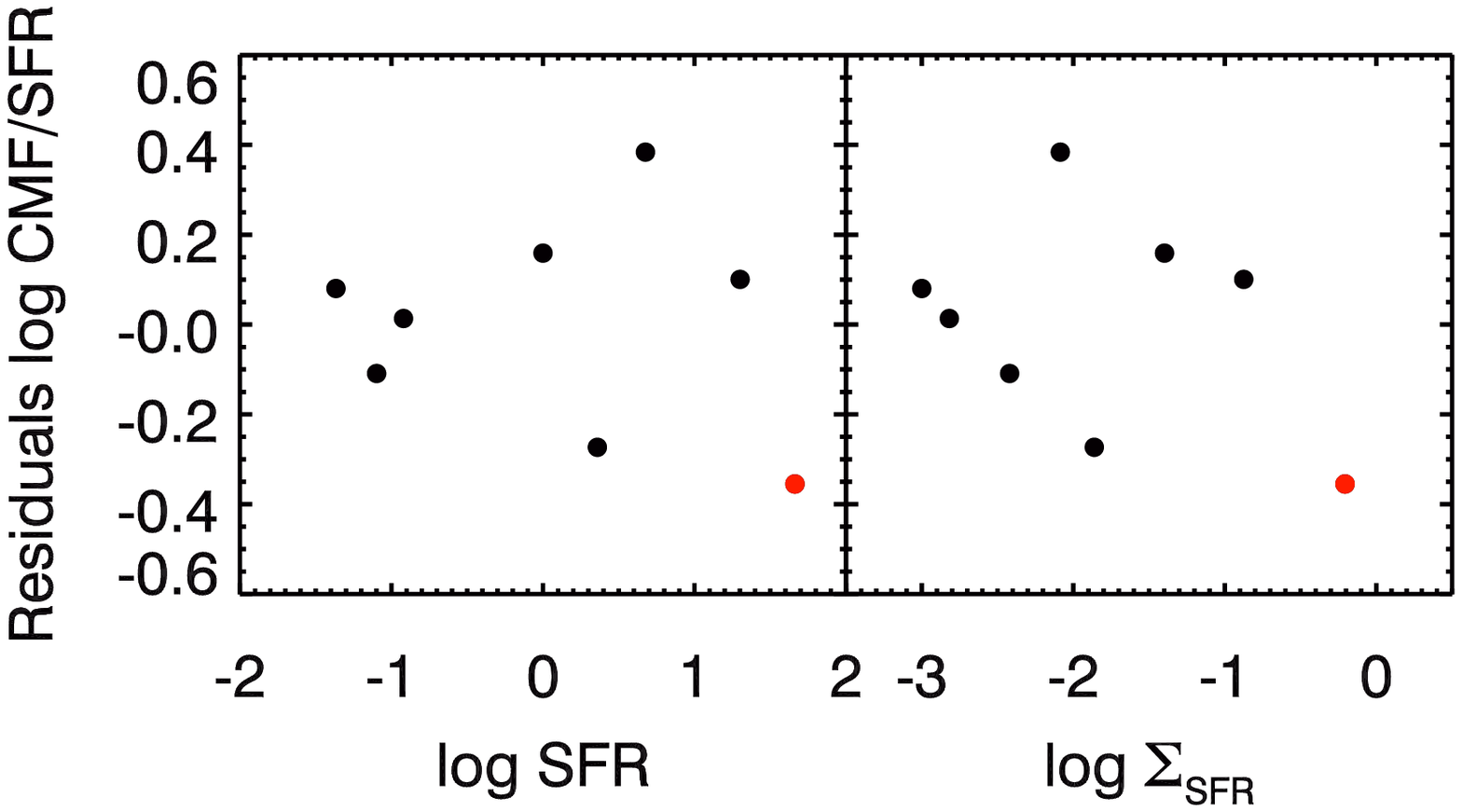}
\vspace{9mm}
\caption{Residuals of CMF/SFR with respect to log(SFR) and
  log($\Sigma_{\rm{SFR}}$), similar to Figure 4 of CFW15. The red
  point on both plots is for NGC 3256. No systematic trends are found.}
\label{residuals}
\end{center}
\end{figure}

\begin{enumerate}

\item The LF follows a power law with index $\alpha =
  -2.23 \pm 0.07$ where $m_V < 21.5$ for the sample combining inner
  (with $m_V < 21.5$) and outer (with $m_V < 23.0$) regions of NGC
  3256. Measuring $\alpha$ for the inner and outer regions separately
  yielded $\alpha \approx -2.1$. These values agree with previous work.

\item We found that the age distribution can be described by a power law
  with index $\gamma \approx -0.67 \pm 0.08$ for independent mass ranges and
  for catalogs from the inner and outer areas of the galaxy
  separately. These values can be interpreted as a destruction rate of $\sim
  80\%$ each decade in time and are consistent with typical values of
  $\gamma = -0.8 \pm 0.2$ found in other systems.

\item The mass functions in various cluster age ranges are well
  described by a power law with index $\beta$. Young
  (log$(\tau/\rm{yr}) < 7$) clusters follow a robust $\beta =
  -1.86 \pm 0.34$. We found that $7 <$ log$(\tau/\rm{yr}) < 8$
  clusters are better described by $\beta = -1.31 \pm 0.36$, while $8
  <$ log$(\tau/\rm{yr}) < 8.6$ clusters follow $\beta = -2.08 \pm
  0.45$. We investigated a number of sources of uncertainty in $\beta$
  and found that uncertainties agree with the formal uncertainties
  given in Figure \ref{dndm}.

\item In order to test for the effect that image resolution can have
  on cluster properties, we artificially degraded an image of the
  Antennae and created independent catalogs from the degraded and original
  images. While less than half of the image-smoothed sources were
  detected in the original image, the LFs, color distributions, and
  age distributions produced from both catalogs were very similar. We
  conclude that reliable measurement of the ages and luminosities of
  star clusters is not significantly hampered by the distance to NGC 3256.

\item We considered two different methods that measure the efficiency
  with which clusters form in a galaxy. The CMF/SFR statistic was
  measured for eight galaxies, including NGC 3256, and had a
  dispersion of $\sigma$(log $A) = 0.24$. We measured $\Gamma$ and
  found a value of $33\%$ from clusters younger than 10 Myr, and we
  discussed the different parameters and assumptions that affect this method.

\end{enumerate}

We thank the referee for helpful comments. R.C. is grateful
for support from NSF through CAREER award 0847467.
This work is based on observations made with the NASA/ESA Hubble Space
Telescope, and obtained from the Hubble Legacy Archive, which is a collaboration
between the Space Telescope Science Institute (STScI/NASA), the Space
Telescope European Coordinating Facility (ST-ECF/ESA) and the Canadian
Astronomy Data Centre (CADC/NRC/CSA). This work was supported in part
by NASA through grant GO-9735-01 from the Space Telescope Science
Institute, which is operated by AURA, INC, under NASA contract
NAS5-26555. This research has made use of the NASA/IPAC Extragalactic
Database, which is operated by the Jet Propulsion Laboratory,
California Institute of Technology, under contract with NASA.

\pagebreak

\pagebreak

\pagebreak

\begin{table}
\caption{Catalog of candidate star clusters in NGC 3256. Apparent
  magnitudes ($m_{\lambda}$) and photometric errors
  ($\sigma_{\lambda}$) are listed, as well as best-fit cluster ages in
  log years ($\tau$), masses ($M$) in log$(M/M_{\odot})$, and internal
  reddening ($E(B-V)$) as calculated from the SED fit. (This table is
  available in its entirety in machine-readable form.)}
\label{catalog}
\begin{center}
\begin{tabular}{lccccccccccccccc}
\hline\hline
ID & R.A. & Decl. & $m_U$ & $\sigma_U$ & $m_B$ & $\sigma_B$ & $m_V$ &
$\sigma_V$ & $m_{H\alpha}$ & $\sigma_{H\alpha}$ & $m_I$ & $\sigma_I$ & $M$ & $\tau$ & $E(B-V)$ \\
\hline

1 & 10 27 52.352 & -43 54 33.30 & 22.87 &  0.08 & 23.12 &  0.02 &
22.81 &  0.01 & 25.45 &  0.04 & 22.36 &  0.02 & 5.25 & 8.31 & 0.00 \\
2 & 10 27 52.377 & -43 54 32.71 & 23.16 &  0.11 & 23.13 &  0.02 &
22.90 &  0.01 & 25.54 &  0.05 & 22.30 &  0.02 & 5.25 & 8.41 & 0.00 \\
3 & 10 27 49.477 & -43 54 31.89 & 20.70 &  0.02 & 22.15 &  0.03 &
22.25 &  0.03 & 25.29 &  0.11 & 22.07 &  0.04 & 4.35 & 6.82 & 0.00 \\
4 & 10 27 49.490 & -43 54 31.77 & 20.58 &  0.03 & 21.81 &  0.04 &
21.79 &  0.03 & 24.75 &  0.09 & 21.56 &  0.04 & 4.56 & 6.84 & 0.00 \\
5 & 10 27 49.626 & -43 54 31.75 & 21.25 &  0.03 & 22.73 &  0.03 &
22.65 &  0.03 & 24.01 &  0.06 & 22.34 &  0.04 & 4.16 & 6.74 & 0.04 \\
6 & 10 27 49.491 & -43 54 31.44 & 19.40 &  0.01 & 20.81 &  0.02 &
20.87 &  0.02 & 23.89 &  0.04 & 20.58 &  0.02 & 4.93 & 6.84 & 0.00 \\
7 & 10 27 49.724 & -43 54 31.12 & 20.52 &  0.02 & 21.61 &  0.01 &
21.38 &  0.01 & 23.06 &  0.07 & 20.82 &  0.02 & 4.89 & 6.76 & 0.22 \\
8 & 10 27 49.723 & -43 54 31.00 & 20.33 &  0.01 & 21.53 &  0.02 &
21.35 &  0.02 & 23.24 &  0.11 & 20.76 &  0.02 & 4.95 & 6.86 & 0.14 \\
9 & 10 27 50.821 & -43 54 29.85 & 22.08 &  0.05 & 22.92 &  0.02 &
22.55 &  0.02 & 24.67 &  0.10 & 21.90 &  0.03 & 5.25 & 7.76 & 0.16 \\
10 & 10 27 50.805 & -43 54 29.81 & 22.70 &  0.10 & 23.37 &  0.05 &
22.75 &  0.04 & 25.47 &  0.33 & 21.97 &  0.05 & 5.46 & 8.01 & 0.28 \\

\hline
\end{tabular}
\end{center}
\end{table}

\begin{table}
\caption{Different estimates of the fraction of stars forming in
  clusters ($\Gamma$) in NGC 3256 using different assumptions. Column
  4 shows $\Gamma$ with the given assumptions, while column 5 shows
  $\Gamma$ values when the three additional 25\% correction factors
  used by G10 have been applied (see Section \ref{Gamma} for details).}
\label{gammas}
\begin{center}
\begin{tabular}{ccc|cc}
\hline\hline
$\beta$ & SFR ($M_{\odot}$ yr$^{-1}$) & $\Delta t$ (Myr) & $\Gamma$ & $\Gamma'$\\
\hline
-1.8 & 46 & 7 & 0.24 & 0.47\\
-1.8 & 46 & 10 & 0.17 & 0.33\\
-1.8 & 50 & 7 & 0.22 & 0.43\\
-1.8 & 50 & 10 & 0.15 & 0.30\\
-1.8 & 85 & 7 & 0.13 & 0.25\\
-1.8 & 85 & 10 &  0.09 & 0.18\\
-2.0 & 46 & 7 & 0.36 & 0.71\\
-2.0 & 46 & 10 & 0.25 & 0.50\\
-2.0 & 50 & 7 & 0.33 & 0.65\\
-2.0 & 50 & 10 & 0.23 & 0.46\\
-2.0 & 85 & 7 & 0.20 & 0.38\\
-2.0 & 85 & 10 & 0.14 & 0.27\\
-2.2 & 46 & 7 & 0.98 & 1.91\\
-2.2 & 46 & 10 & 0.68 & 1.33\\
-2.2 & 50 & 7 & 0.90 & 1.75\\
-2.2 & 50 & 10 & 0.63 & 1.23\\
-2.2 & 85 & 7 & 0.53 & 1.03\\
-2.2 & 85 & 10 & 0.37 & 0.72\\
\hline
\end{tabular}
\end{center}
\end{table}

\pagebreak

\end{document}